\documentclass{nature_wf2}
\usepackage{lineno}
\usepackage{multirow}
\usepackage{amsmath}
\usepackage{amssymb}
\usepackage{graphicx}
\usepackage{lineno}
\title{Transition to a Moist Greenhouse with CO$_2$ and solar forcing}

\author{Max Popp$^{1,2 \star}$, Hauke Schmidt$^1$ \& Jochem Marotzke$^1$}

\renewcommand{\baselinestretch}{2.} 
\begin{document}
%\linenumbers
\maketitle

\begin{affiliations}
 \item Max Planck Institute for Meteorology, Bundesstr. 53, 20146 Hamburg, Germany 
 %\item Geophysical Fluid Dynamics Laboratory, 201 Forrestal Road,  Princeton, NJ 08540-6649, USA
 \item Program in Atmospheric and Oceanic Sciences, Princeton University, 300 Forrestal Road, Sayre Hall, Princeton, NJ 08544, USA; located at: NOAA / Geophysical Fluid Dynamics Laboratory, Princeton, NJ, USA \\
  $\star$ email: mpopp@princeton.edu
%\item Put institutions in this environment and
 %\item separate with \verb|\item| commands.
\end{affiliations}

\begin{abstract}
Water-rich planets such as Earth are expected to become eventually uninhabitable, because liquid water does not remain stable at the surface as surface temperatures increase with the solar luminosity over time. Whether a large increase of atmospheric concentrations of greenhouse gases such as CO$_2$ could also destroy the habitability of water-rich planets has remained unclear. We show with three-dimensional aqua-planet numerical experiments that CO$_2$-induced forcing as readily destabilizes the climate as does solar forcing. The climate instability is caused by a positive cloud feedback. The climate does not run away, but instead attains a new steady state with global-mean sea-surface temperatures above 330 K. The upper atmosphere is considerably moister in this warm steady state than in the reference climate, implying that the planet would be subject to substantial loss of water to space. For either a certain range of elevated CO$_2$ concentrations or solar irradiation, we find both cold and warm equilibrium states, implying that the transition to the warm state may not simply be reversed by removing the additional forcing. 
\end{abstract}
%\newpage

Water-rich planets such as Earth lose water by photo-dissociation of water vapour in the upper atmosphere and by the subsequent escape of hydrogen. On present-day Earth the loss occurs very slowly, because the mixing ratio of water vapour in the upper atmosphere is very low. But significant loss of water could occur over geological time-scales if the surface temperature were around 70 K warmer than it is today\cite{kasting_runaway_1988,kasting_habitable_1993,kopparapu_habitable_2013}. For these high surface temperatures, the tropopause is expected to climb to high altitudes. As a consequence, the cold-trapping of water vapour at the tropopause becomes ineffective, because the mixing ratio of water vapour increases with the rising tropopause. Steady states in which the mixing ratio in the upper atmosphere is sufficiently high for a water-rich planet to lose most of its water inventory in its life-time are known as Moist-Greenhouse states\cite{kasting_response_1984}. A planet in this state would eventually become uninhabitable as all water is lost to space. For an Earth-like planet around a sun-like star, a Moist Greenhouse would be attained if the mixing ratio in the upper atmosphere exceeds around 0.1 $\%$\cite{kasting_runaway_1988}. For comparison, the mixing ratio in Earth's stratosphere is presently around two orders of magnitude smaller. \par
Moist Greenhouse states were found and described in several studies with one-dimensional models\cite{kasting_runaway_1988,kasting_habitable_1993,kopparapu_habitable_2013,wordsworth_water_2013,ramirez_can_2014} and have recently been found for terrestrial planets with three-dimensional models in different setups\cite{abe_habitable_2011,yang_stabilizing_2013,wolf_evolution_2015}. However, not all three-dimensional studies found stable Moist Greenhouse states\cite{ishiwatari_numerical_2002,ishiwatari_dependence_2007,leconte_increased_2013}.  Instead the climate of these models would destabilize into a Runaway Greenhouse, a self-reinforcing water-vapour feedback-loop, before the Moist Greenhouse is attained. A few studies applied large forcings but the employed models became numerically unstable before the Moist Greenhouse regime was attained\cite{boer_climate_2005,heinemann_warm_2009,wolf_delayed_2014}. Therefore, it remains unclear whether planets would attain a Moist Greenhouse state before a Runaway Greenhouse occurs, especially for planets on an Earth-like orbit, where the only two previous studies with state-of-the-art general circulation models (GCM) gave contradicting results\cite{leconte_increased_2013,wolf_evolution_2015}. Moreover, all three-dimensional studies investigating Moist-Greenhouse states only applied solar forcing without considering greenhouse-gas forcing. Several studies have applied strong greenhouse-gas forcing, but either did not run their simulations to sufficiently high temperatures\cite{hansen_efficacy_2005,meraner_robust_2013,wolf_hospitable_2013,bloch_feedback_2015} or did not investigate the emergence of a Moist Greenhouse\cite{russell_fast_2013}.   
Greenhouse-gas forcing has long been assumed to be ineffective at causing Moist Greenhouse states, because the greenhouse effect of any additional greenhouse gas would eventually be rendered ineffective by the increasing greenhouse effect of water vapour with increasing temperatures. Furthermore, large greenhouse-gas forcings would lead to a cooling of the upper atmosphere, which would push the Moist Greenhouse limit to much higher surface temperatures\cite{wordsworth_water_2013}. However, if clouds are considered, these arguments may not apply, because clouds themselves can contribute to the climate becoming unstable\cite{popp_initiation_2015}. \par 
Here we compare for the first time with a state-of-the-art GCM, namely ECHAM6\cite{stevens_atmospheric_2013}, how effective solar and CO$_2$ forcing are at causing a transition to a Moist Greenhouse. We couple the atmosphere to a slab ocean and choose an aqua-planet setup (fully water-covered planet) in perpetual equinox. This idealized framework is better suited than a present-day Earth setting to understand the involved dynamics while preserving the major feedback mechanisms of Earth\cite{medeiros_using_2015}. It also avoids conceptual problems with the representation of land-surface processes at high temperatures. We turn off sea ice in order to investigate the possibility of solely cloud-induced multiple steady states that were recently found in a one-dimensional study\cite{popp_initiation_2015}. We modify the model such that it can deal with surface temperatures of up to 350 K (see Methods). 

\section*{Results}
\begin{figure}
  \hspace{2pc}
  \vspace{0pc}
  \noindent\includegraphics[width=35pc,angle=0]{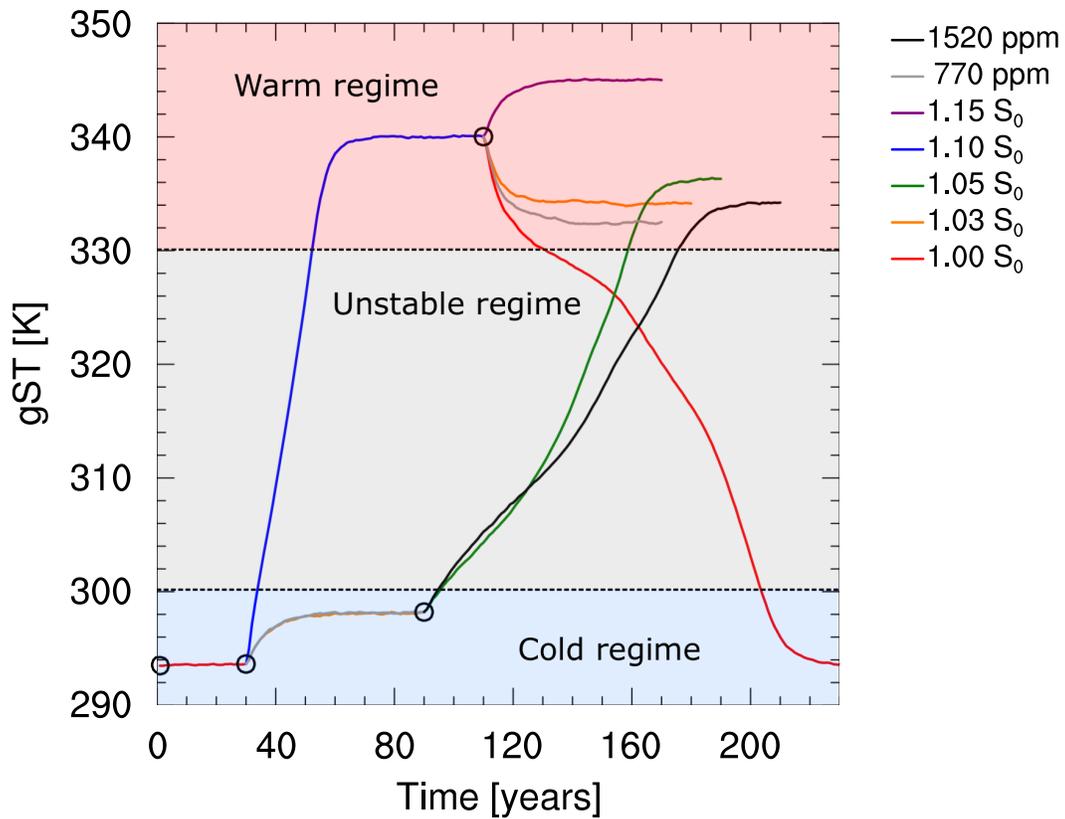}\\
  \caption{\label{TIME_SST} $\mid$ \textbf{Temporal evolution of gST.} The circles denote the four states from which new simulations are started. For both a TSI of 1.00 $S_0$ and 1.03 $S_0$ as well as for a CO$_2$ concentration of 770 ppm, two simulations with different initial conditions are performed. The lines are interpolated from annual means.}
\end{figure}

\begin{figure}
  \hspace{-2pc}
  \noindent\includegraphics[width=42pc,angle=0]{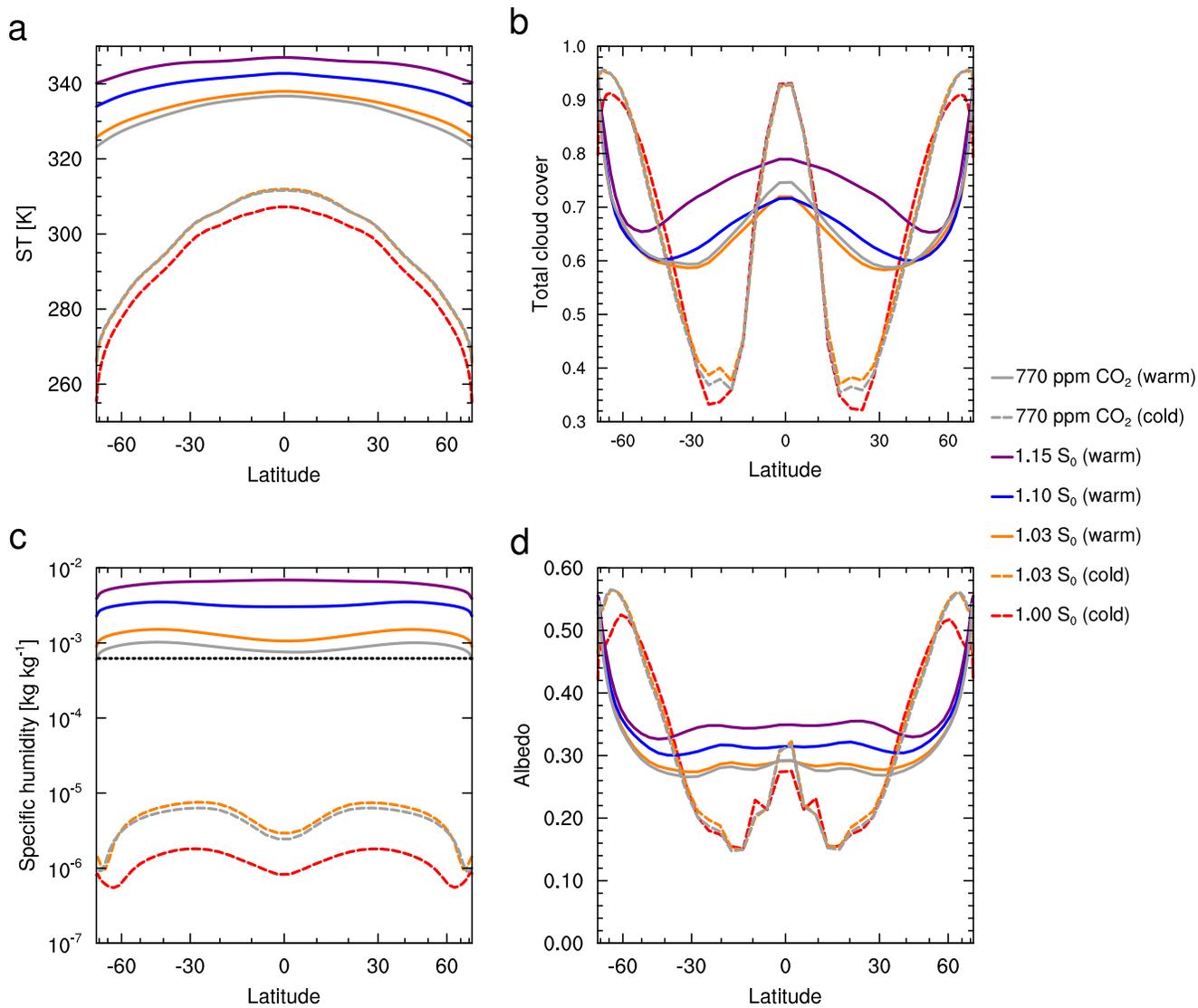}\\
  \caption{\label{30y_LAT} $\mid$ \textbf{Zonal means in steady state.} Panel a) shows the surface temperature (ST), panel b) the total cloud cover, panel c) the specific humidity at the top level and panel d) the effective albedo. The effective albedo is defined as the ratio of the zonal and temporal means of the reflected solar radiation divided by the zonal and temporal means of the incoming solar radiation. The black horizontal line in panel c) indicates the Moist-Greenhouse limit\cite{kasting_runaway_1988}. The temporal mean is taken over a period of 30 years. The horizontal axes are scaled with the cosine of the latitude.}
\end{figure}

To assess the dependence of the climate state on total solar irradiance (TSI) for fixed CO$_2$ levels (at 354 ppm volume mixing ratio), we apply a total of five different TSI-values that range from the present-day value on Earth (S$_0$) to 1.15 times that value. We find two regimes of steady states that are separated by a range of global-mean surface temperature (gST) for which stable steady states are not found (Fig. \ref{TIME_SST}).  The regime of steady states with  gST of up to around 298 K exhibits similar features as present-day Earth climate, such as a large pole-to-equator surface-temperature contrast (Fig. \ref{30y_LAT}a) and a similar meridional distribution of cloud cover (Fig. \ref{30y_LAT}b). Hence this regime of steady-state can be considered to be Earth-like. In contrast, the warm regime of steady states with  gST above 334 K is characterized by a considerably smaller pole-to-equator surface-temperature difference and a substantially different meridional distribution of cloud cover (Fig. \ref{30y_LAT}a,b). This illustrates that the dynamics in the warm regime are quite different from the present-day Earth regime. Most importantly, the mixing ratio of water vapour at the uppermost levels is considerably higher in the warm regime and exceeds the Moist-Greenhouse limit (Fig. \ref{30y_LAT}c). Therefore, a planet in such a state would be losing water to space at a fast rate. The minimum TSI required to cause a climate transition from the Earth-like to the warm regime lies between 1.03 S$_0$ and 1.05 S$_0$, whereas the maximum TSI to cause a climate transition from the warm back to the Earth-like regime lies between 1.00 S$_0$ and 1.03 S$_0$ (Fig. \ref{TIME_SST}). Consequently, there are two different stable steady states for a TSI of 1.03 S$_0$. Since sea ice is turned off in our model, this double steady state is entirely a consequence of atmospheric processes. In the warm regime, the cloud-albedo increases at all latitudes with TSI, thus providing an efficient way to stabilize the climate against increased radiative forcing (Fig. \ref{30y_LAT}d). \par 
  
\begin{figure}
  \hspace{2pc}
  \vspace{-1pc}
  \noindent\includegraphics[width=35pc,angle=0]{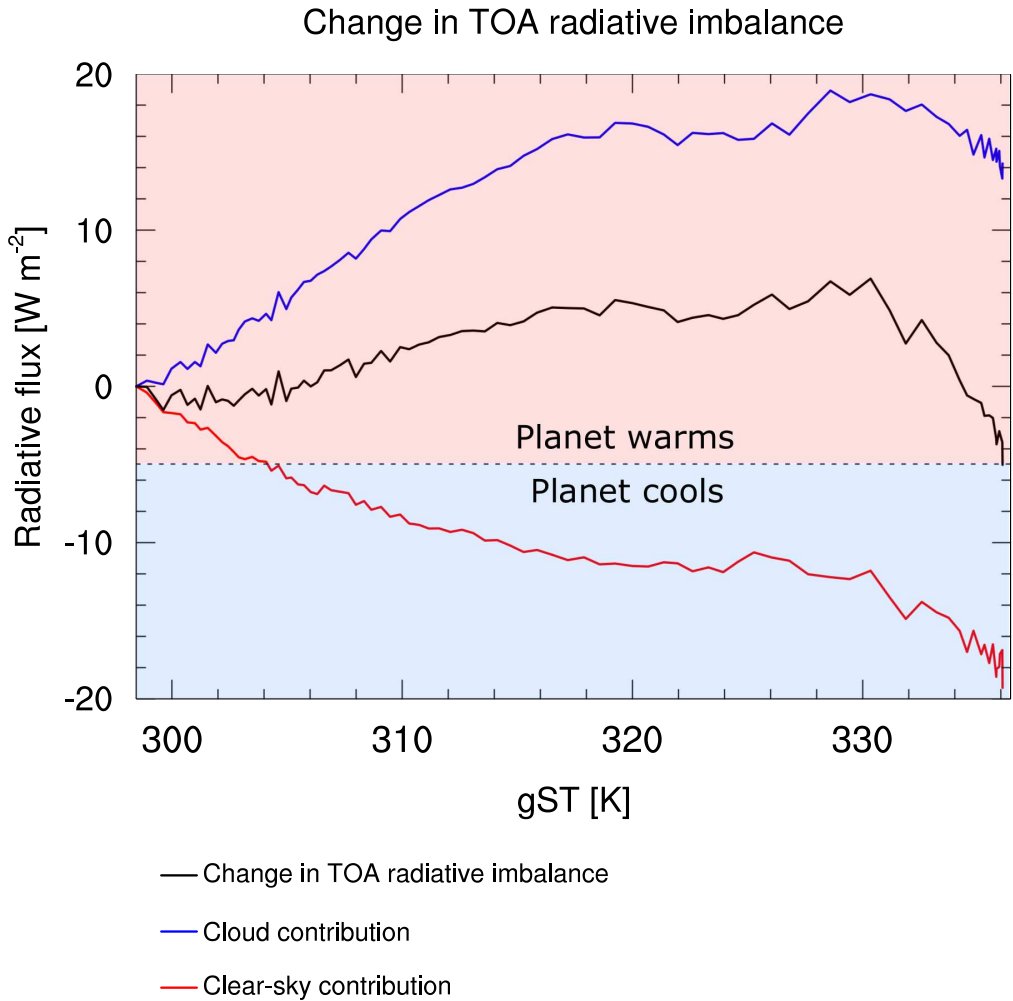}\\
  \caption{\label{transient} $\mid$ \textbf{Change of TOA radiative imbalance as a function of gST.} The black line shows the change in global and annual mean of the net TOA radiative imbalance, the blue line shows the cloud-radiative contribution and the red line the clear-sky contribution from the transient simulation with a TSI of 1.05 S$_0$. The cloud-radiative and clear-sky contributions sum up to the net TOA radiative imbalance. The changes are calculated by subtracting the global and annual means over the first year of the respective quantities. The horizontal line corresponds to the negative value of the initial TOA imbalance. Hence, for a steady state to be attained, the TOA radiative imbalance must touch or intersect the horizontal line.}
\end{figure}

In order to understand the processes governing the climate transition from the cold to the warm regime, we focus now on the transient simulation with a TSI of 1.05 S$_0$. The climate instability is evidenced by an increase in planetary energy uptake with increasing gST for gST between 300 K and 330 K (Fig. \ref{transient}). The energy uptake is diagnosed from the top-of-the-atmosphere (TOA) radiative imbalance, counting positive downward. The instability is caused by the cloud-radiative contribution to the energy uptake that increases with increasing gST for gSTs below 330 K. The clear-sky contribution is decreasing with increasing gST and does thus not contribute to the climate instability. At gST above 330 K, the cloud-radiative contribution decreases again with increasing gST. This allows together with the clear-sky contribution to attain a new steady state. Thus, clouds destabilize the climate at lower gST and then stabilize again at higher gST.  \par
\begin{figure}
	\hspace{-3pc}
	\noindent\includegraphics[width=45pc,angle=0]{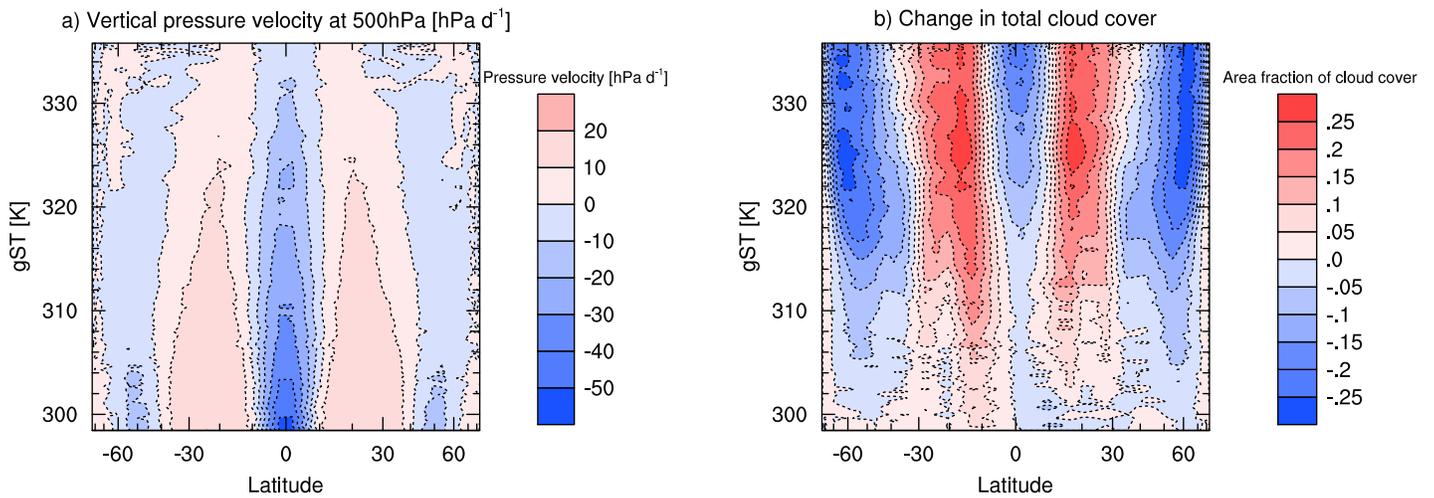}\\
	\caption{\label{dyn_ch} $\mid$ \textbf{Large-scale circulation and cloud cover during the climate transition.} Panel a) shows the zonal and annual mean of the vertical pressure velocity as a function of latitude and gST for the transient period of the simulation with a TSI of 1.05 S$_0$. Panel b) shows the same as panel a) but for the change in zonal and annual mean of cloud cover from the first year of simulation.}
\end{figure}
\begin{figure}
	\hspace{-1pc}
	\vspace{-0pc}
	\noindent\includegraphics[width=40pc,angle=0]{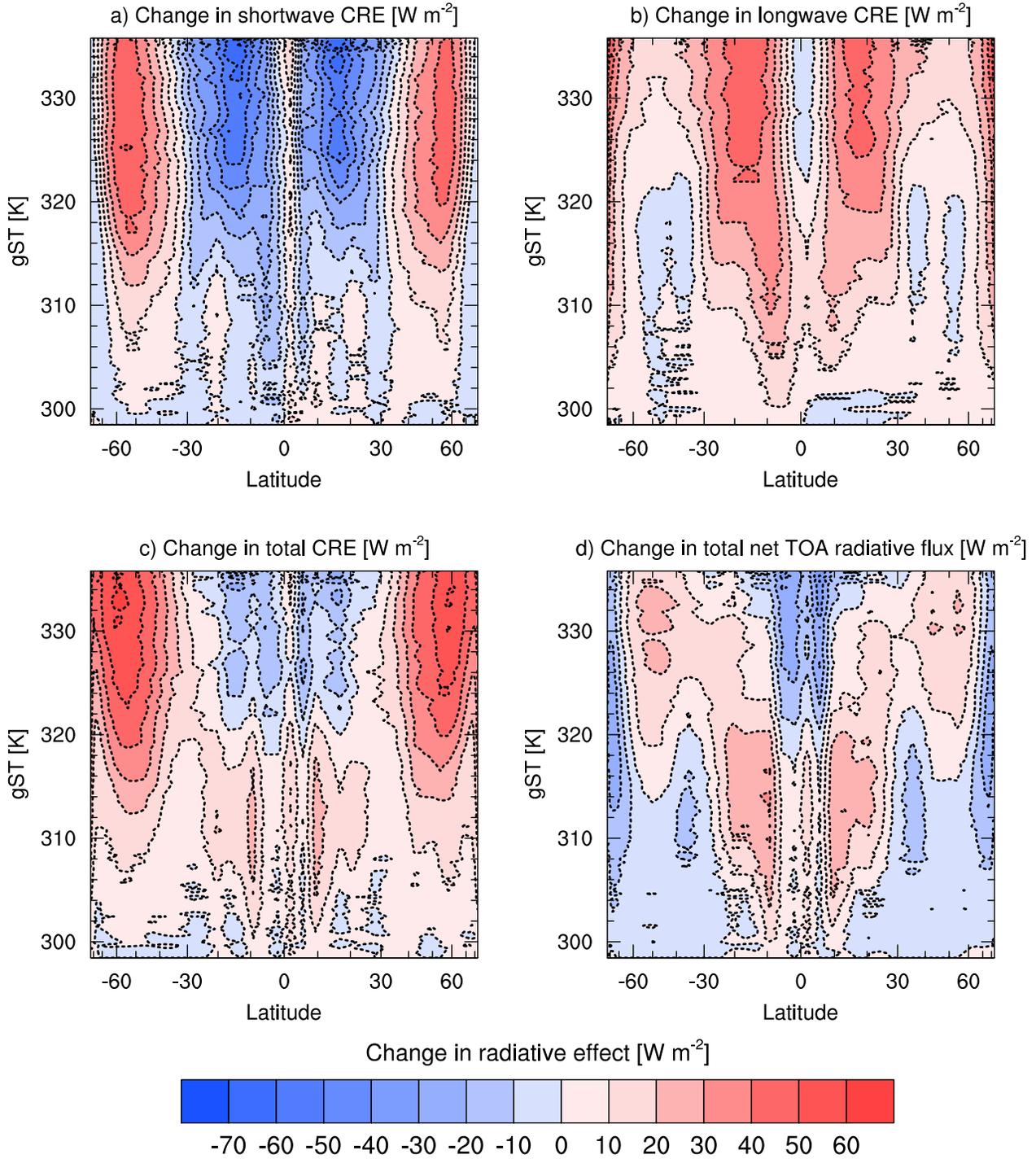}\\
	\caption{\label{cre_ch} $\mid$ \textbf{Zonal means of changes in radiative effects during the climate transition.} Panel a) shows the change of zonal and annual mean of the shortwave CRE, panel b) the change of the longwave CRE, panel c) the change of the longwave CRE and panel d) the change of the TOA radiative imbalance. All panels use the same colour bar. The horizontal axes are scaled with the cosine of the latitude.}
\end{figure}
\begin{figure}
	\hspace{1pc}
	\vspace{-2pc}
	\noindent\includegraphics[width=35pc,angle=0]{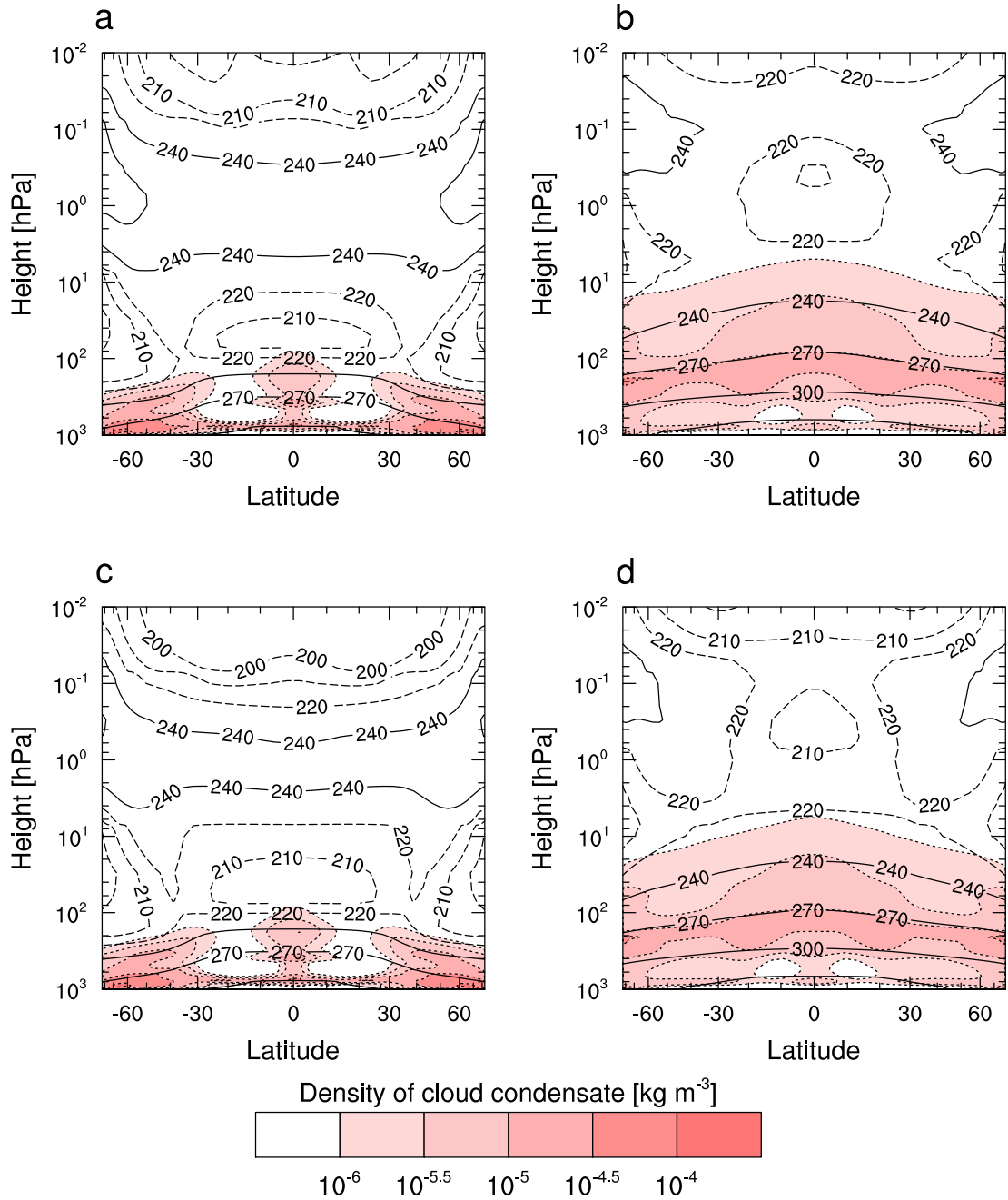}\\
	\caption{\label{condensate} $\mid$ \textbf{Zonal means of cloud condensate in steady state.} Panel a) shows the temporal mean over the last 30 years of simulation of the cloud condensate for a TSI of 1.03 S$_0$ in the cold regime and panel b) shows the same quantity for the same TSI but in the warm regime. Panel c) shows the same quantity obtained with a TSI of 1.00 S$_0$ but with atmospheric CO$_2$ concentrations of 770 ppm in the cold regime and panel d) with the same CO$_2$ concentration and TSI in the warm regime. The contours denote temperatures in Kelvin, with the solid lines denoting the contours for 240 K, 270 K, 300 K and 330 K and the dashed lines for the contours of 200 K, 210 K and 220 K. The vertical axes are the height in terms of pressure of dry air and the horizontal axes are the latitudes scaled with their cosines.}
\end{figure}

The changes in cloud-radiative contribution to the energy uptake (henceforth simply referred to as cloud-radiative effects, CRE) are caused by the weakening of the large-scale circulation (Fig. \ref{dyn_ch}a) and the increase of water vapour in the atmosphere with increasing gST. The weakening of the circulation causes tropical convection to spread more evenly around the tropics. As a consequence, deep convective clouds with low cloud-top temperatures become more frequent in the subsidence region of the Hadley circulation (Fig. \ref{dyn_ch}b). This in turn leads to a very strong increase in longwave CRE in this region, which dominates the increase in shortwave CRE (Fig. \ref{cre_ch}a,b). However, as the gST increases further and the specific humidity in the atmosphere increases, the clouds become thicker and thus more reflective, whereas the longwave CRE does not increase as fast anymore. Therefore, the total CRE decreases again with increasing gST which contributes to the stabilization of the climate at gST above 330 K (Fig. \ref{cre_ch}c). In general, the changes in total CRE dominate the changes in clear-sky radiative effect in the tropics (Fig. \ref{cre_ch}c,d). In the extra-tropics, the weakening of the large-scale circulation leads to a steady decrease in cloud cover everywhere except at very high latitude (Fig. \ref{dyn_ch}a,b). Therefore, the shortwave CRE increases in the extra-tropics (Fig. \ref{cre_ch}a). Since the tropopause deepens with increasing surface temperatures (not shown), the difference between the temperature at the surface and at the cloud tops increases as well and leads also to an increase in longwave CRE despite the decrease in cloud cover (Fig. \ref{cre_ch}b). Whereas the changes in the clear-sky radiative effect dominate the changes in CRE in most of the extra-tropics for gST up to 315 K, the changes in CRE increasingly dominate the extra-tropical response at gST above. This supports the idea that at high gST changes in CRE are more important than changes in clear-sky radiative effect and dominate the climate response. In general the weak large-scale circulation in the warm regime leads to a much more uniform meridional distribution of cloud condensate at all levels than in the cold regime (Figure \ref{condensate}). \par

% We turn now to the simulations with 1.00 S$_0$ but different CO$_2$-concentrations.CO$_2$-concentrations are first increased to 770 ppm, which corresponds to an equivalent adjusted forcing as is caused by an increase from 1.00 S$_0$ to 1.03 S$_0$. The adjusted forcing is defined to be the temporal and global means of the energy uptake over the first year of simulation.

We start our comparison of CO$_2$-induced to solar forcing by increasing CO$_2$ concentrations to 770 ppm while keeping the TSI fixed to 1.00 S$_0$. This corresponds to an equivalent adjusted forcing as is caused by an increase of TSI from 1.00 S$_0$ to 1.03 S$_0$. The adjusted forcing is defined to be the temporal and global mean of the energy uptake over the first year of simulation. The results suggest that the increase of the CO$_2$ concentrations leads to an equivalent warming and a similar meridional distribution of surface temperatures and clouds as the increase in TSI does (Fig. \ref{30y_LAT}a,d). Starting from the final state of the simulation with a TSI of 1.03 S$_0$, we then increase the CO$_2$ concentrations to 1520 ppm and set the TSI back to 1.00 S$_0$ (Fig. \ref{TIME_SST}). The combined effect leads to an adjusted forcing that is equivalent to increasing the TSI to 1.05 S$_0$. In this case the aqua-planet undergoes a climate transition into the warm regime (Fig. \ref{TIME_SST}). Thus, the aqua-planet can as readily be forced to transition from the Earth-like to the warm regime by increasing CO$_2$ concentrations as by increasing the TSI. When starting in the warm regime, a reduction of CO$_2$ concentration to 770 ppm does not cause the planet to fall back into the Earth-like regime, but the aqua-planet remains in the warm regime. Therefore, the aqua-planet also exhibits a bistability of the climate for a TSI of 1.00 S$_0$ and a CO$_2$ concentration of 770 ppm. \par
Overall, the results suggest that the aqua-planet behaves similarly for solar forcing and CO$_2$-induced forcing (Fig. \ref{TIME_SST}, \ref{30y_LAT}, \ref{condensate}). The most notable difference is that the steady-state gST in the warm regime is around two degrees lower for CO$_2$-induced forcing. The likely reason for this is that the thermal absorption by water vapour overlaps with the thermal absorption by CO$_2$ in the warm moist atmosphere, which renders the greenhouse effect of CO$_2$ less effective. However, since the climate instability in our simulations is caused by cloud-radiative effects at a gST at which the atmosphere is not yet sufficiently opaque to cancel the greenhouse effect of CO$_2$, CO$_2$-induced forcing can as easily cause a climate transition to the Moist Greenhouse as solar forcing does.  \par

\section*{Discussion}
A recent study using the same model but in a different version found that Earth's climate remains stable for CO$_2$ concentrations of at least 4480 ppm\cite{meraner_robust_2013}, whereas our study suggests that such concentrations would lead to a climate transition. Studies of Earth with other GCMs also found the climate to remain stable for higher CO$_2$ concentrations than we do\cite{hansen_efficacy_2005,wolf_hospitable_2013}. However, the initial climate of our aqua-planet is around 6 K warmer than the one of present-day Earth. Such a warming would be attained by a quadrupling of CO$_2$ in the different version of our model used in ref. 17. By a simple estimate, this other study would thus have explored CO$_2$ concentrations of up to a fourth of 4480 ppm, hence 1120 ppm, if the simulations were started from a climate similar to ours. Therefore, if we account for the difference in the initial climates, the results of the two studies are not in contradiction. Indeed, the climate of the model version used in ref. 17 was recently shown to become unstable when the CO$_2$ concentrations were increased from 4480 ppm to 8960 ppm (eventually leading to numerical failure of their model)\cite{bloch_feedback_2015}. Nonetheless, the forcing required to cause a climate transition would certainly be higher on present-day Earth than on our aqua-planet, even with our version of the model. Several other studies of Earth have found lower climate sensitivities to relatively large CO$_2$ forcing than we do which supports this notion\cite{hansen_efficacy_2005,wolf_hospitable_2013,russell_fast_2013}.    \par

\begin{figure}
	\hspace{-1pc}
	\vspace{-1pc}
	\noindent\includegraphics[width=40pc,angle=0]{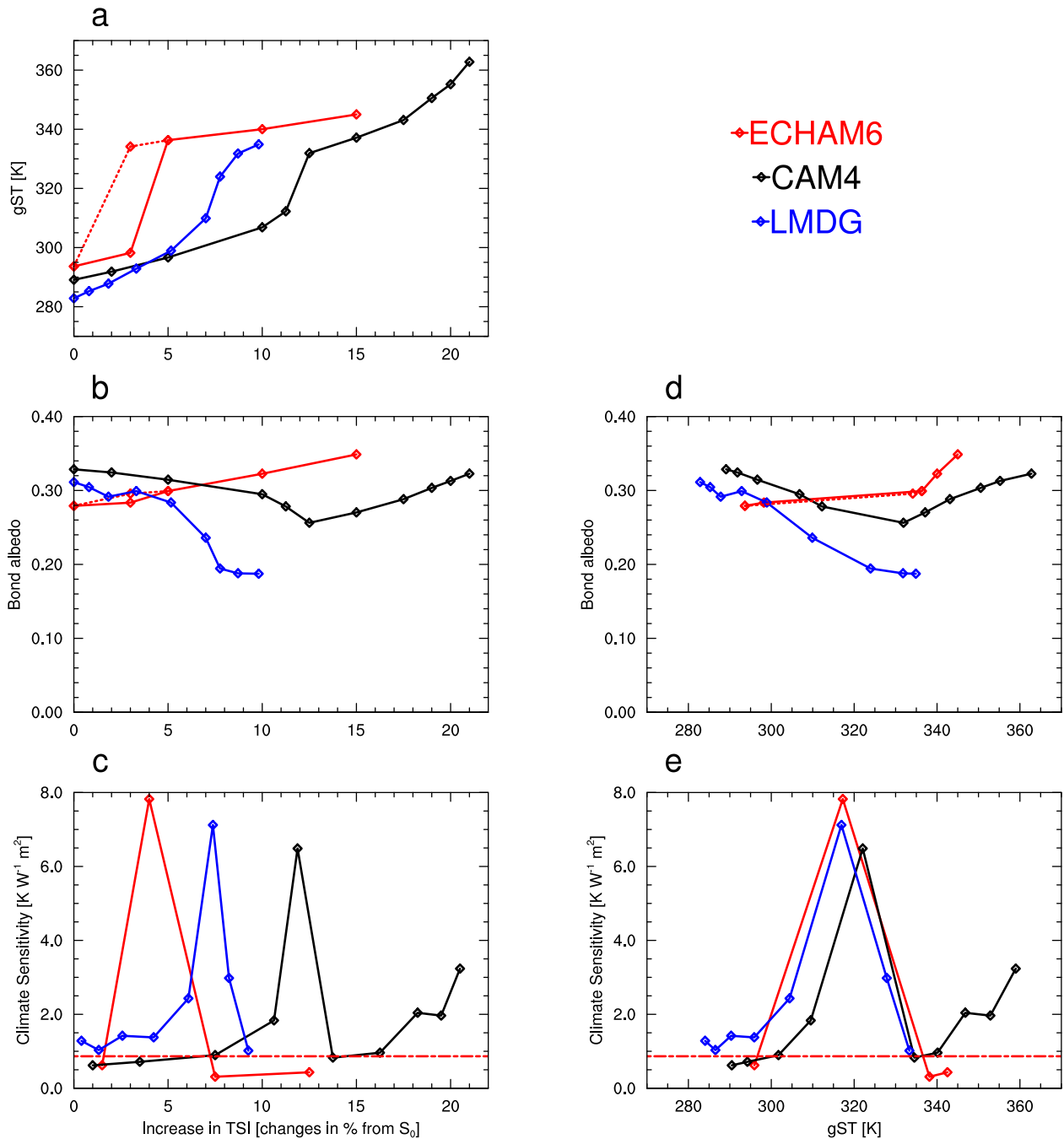}\\
	\caption{\label{comparison} $\mid$ \textbf{Model intercomparison.} Panel a) shows the gST, panel b) the Bond albedo and panel c) the climate sensitivity as a function of the increase in TSI in per cents of S$_0$. The red solid lines and marks denote the results obtained with ECHAM6 when increasing TSI, whereas the dashed lines and mark denote the results obtained with ECHAM6 but when decreasing TSI. The black lines and marks denote the results obtained with CAM4 in ref. 9 and the blue lines and marks denote the results obtained with LMDG in ref. 12. Panel d) shows the Bond albedo and panel e) the climate sensitivity as a function of the gST. Note that the Bond albedo is equal to the global-mean of the effective albedo.}
\end{figure}

\begin{figure}
	\hspace{0pc}
	\noindent\includegraphics[width=40pc,angle=0]{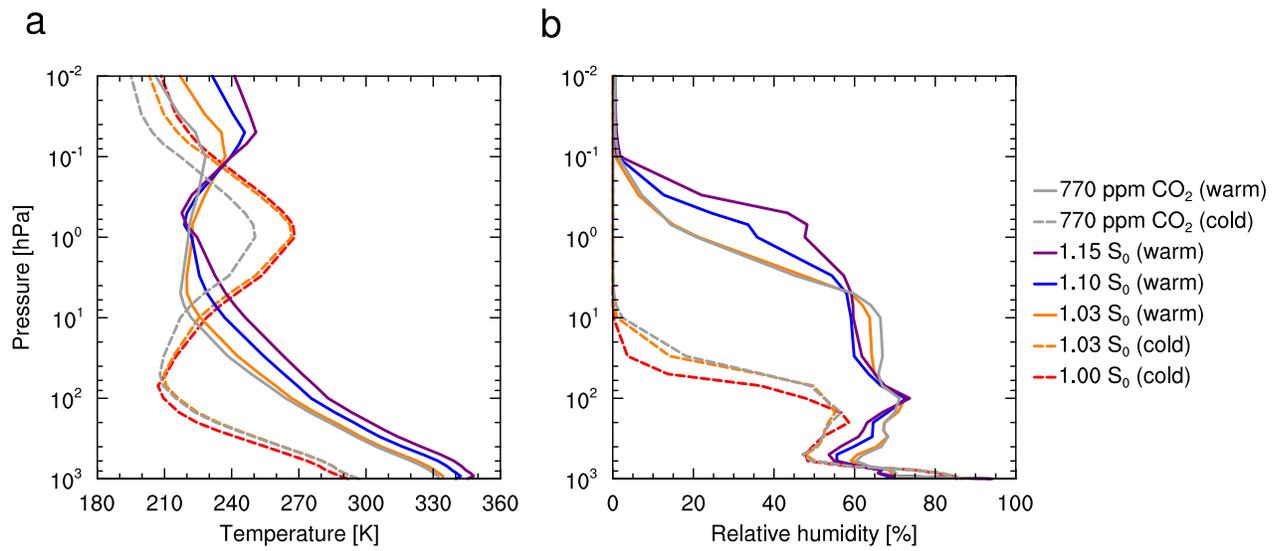}\\
	\caption{\label{profiles} $\mid$ \textbf{Atmospheric profiles of temperature and relative humidity.} Panel a) shows the global-mean vertical profiles of gST and panel b) of relative humidity for the different steady states. The vertical axes are the height in terms of pressure of dry air.}
\end{figure}

Two studies recently investigated climates at gST above 330 K with state-of-the-art GCMs for Earth-like planets\cite{leconte_increased_2013,wolf_evolution_2015}. Ref 9 used a modified version of the Community Atmosphere Model version 4 (CAM4) and ref. 12 a modified version of the Laboratoire de M\'{e}t\'{e}orologie Dynamique Generic (LMDG) climate model to investigate the climate response to strong solar forcing and both studies also found a region of increased climate sensitivity\cite{leconte_increased_2013,wolf_evolution_2015}. Therefore, a region of gST with increased climate sensitivity surrounded by regions of lower climate sensitivity appears to be a robust result, despite some differences in the magnitude of the region (Fig. \ref{comparison}c,e). We calculated the climate sensitivity here following ref. 9, which yields a small climate sensitivity in the cold and in the warm regimes because the method uses the instantaneous forcing. If we use the adjusted forcing, the climate sensitivity is considerably larger, because the fast atmospheric adjustments reduce the initial TOA radiative imbalance quickly. Our model encounters the region of high climate sensitivity for smaller values of TSI (Fig. \ref{comparison}a), because our control climate is the warmest and because the climate instability is encountered at around 300 K whereas the region of increased climate sensitivity starts at around 310 K to 315 K in the two other studies. Our steady-state albedo is a monotonically increasing function of both TSI and gST, in contrast to the two other studies where the albedo decreases with both increasing TSI and gST until either the Moist Greenhouse state is attained or a Runaway Greenhouse occurs (Fig. \ref{comparison}b,d). However, in our transition from the cold to the warm regime the albedo decreases as well with increasing gST between 300 K and 320 K. Only one of the two studies found a stabilizing cloud feedback similar to ours and a moist stratosphere at high gST\cite{wolf_evolution_2015}, whereas the other one found that the cloud feedback is rather destabilizing and that the stratosphere remains dry\cite{leconte_increased_2013}. In general, our results of the simulations with increased solar forcing are qualitatively similar to the ones found in ref. 9 in that we find two regimes of steady states, in that the warm regimes have a similar temperature structure (Fig. \ref{profiles}a), in that the cloud albedo increases with gST and most importantly in the existence of a stable Moist Greenhouse regime. Similarly to ref.  9, the troposphere in the warm regime is characterized in our model by a particular radiative-convective equilibrium with a temperature inversion close to the surface, a somewhat drier region above and a more humid region up to the tropopause. A similar structure has also been found and discussed in two one-dimensional studies\cite{wordsworth_water_2013,popp_initiation_2015}. Ref. 9 argues that the change to the aforementioned radiative-convective equilibrium is crucial for the emergence of stable Moist-Greenhouse states, but our results suggest that the weakening of the large-scale circulation is equally important by allowing the radiative convective regime to spread over the entire tropics. This spread of the convective region over a large fraction of the planet, namely the tropics, also explains some of the similarities between the three-dimensional and the one-dimensional models in the warm regime. Compared to ref. 9 the dry region is more humid in our warm regime and in general more humid than in our cold regime (Fig. \ref{profiles}b). Sensitivity experiments reveal that this is not a consequence of the differences in the treatment of ozone and prescribed oceanic heat transport (see supplementary information). Given the different dynamical cores, radiative transfer schemes, convection schemes and cloud schemes, it is, however, remarkable how many similarities the three models share. \par 

One study investigated strong CO$_2$-forcing and gST above 330 K with the Fast Atmosphere-Ocean Model (FAOM) developed at the Goddard Institute for Space Studies and did not find any region of strongly increased climate sensitivity\cite{russell_fast_2013}. The climate sensitivity increases somewhat but not nearly as much as with our or with the other two state-of-the-art models used for solar forcing. As a consequence, considerably higher values of CO$_2$ are required in that study compared to ours to attain a gST of 330 K. Furthermore, the control climate is colder in that study, and it takes around one doubling of CO$_2$ to attain the gST of our control simulation. The humidity at the top level of FAOM remains roughly one order of magnitude below the Moist Greenhouse limit in the warmest steady-states found in ref. 20. This may be partly due to the CO$_2$ cooling of the upper atmosphere, but could also be a consequence of not running the model to sufficiently high temperatures. We perform a sensitivity experiment where we increase CO$_2$ concentrations to 9000 ppm in order to assess whether increased CO$_2$ concentrations could cause a substantial drying of our upper atmosphere (not shown). The upper-atmosphere specific humidity stays, however, well above the Moist-Greenhouse limit also in that case. Some of the differences between FAOM and ECHAM6 may simply be caused by the use of different setups and parametrisations. But FAOM is a simplified model designed for fast computation and uses simplified cloud physics, which may be the cause for the absence of a region of strongly increased climate sensitivity. So whereas our version of ECHAM6 is rather on the low side of CO$_2$ concentrations required to cause the gST to rise above 330 K, FAOM likely is on the high side of the concentrations. \par

To conclude, we have demonstrated with a state-of-the-art climate model that a water-rich planet might lose its habitability as readily by CO$_2$ forcing as by increased solar forcing through a transition to a Moist Greenhouse and the implied long-term loss of hydrogen. We confirm previous results that a region of increased climate sensitivity exists and show that the climate is unstable in our model in that region due to positive cloud feedbacks caused by a weakening of the large-scale circulation. We also demonstrate that there is hysteresis and that once a transition to the Moist Greenhouse regime has occurred, the process may not simply be reversed by removing the additional forcing. 

\section*{Acknowledgements}
We thank Thorsten Mauritsen for a thorough internal review. We thank Eric Wolf for an insightful discussion and for making the data from CAM4 available to us. We thank J\'{e}r\'{e}my Leconte for making the data from LMDG available to us. We thank Dorian Abbot and two anonymous reviewers for very constructive reviews of this manuscript. We thank the Max Planck Society for the Advancement of Science for financial support.

\section*{Author contributions}
J.M., H.S. and M.P. conceived the study. M.P. performed the changes to the model and conducted the modelling work. J.M.,H.S. and M.P. all contributed substantially to the analysis and the discussion. M.P. wrote the manuscript with comments from J.M. and H.S..

\section*{Competing financial interests}
The authors declare no competing financial interests.

\newpage

\begin{methods}
We employ a modified version of the GCM ECHAM6\cite{stevens_atmospheric_2013} in an aqua-planet setting in which the whole surface is covered by a 50 m deep mixed-layer ocean. We run the model with a spectral truncation of T31, which corresponds to a Gaussian grid with a grid-point spacing of 3.75$^{\circ}$. The atmosphere is resolved vertically by 47 layers up to a pressure (of dry air) of 0.01 hPa. The oceanic heat transport is prescribed by a sinusoidal function of latitude. There is no representation of sea ice in our model and as a consequence water may be colder than the freezing temperature. The orbit of the aqua-planet is perfectly circular with a radius of 1 AU. The obliquity of the aqua-planet is 0 $^{\circ}$. For simplicity, a year is set to be 360 days. The rotation velocity of the aqua-planet corresponds to present-day Earth.  \par
If the tropopause climbs, regions with high ozone concentrations may come to lie in the troposphere in the model, because ozone concentrations are prescribed to climatological values. High ozone concentrations could, however, not occur in the presence of tropospheric water vapour concentrations. Therefore, we limit the tropospheric ozone concentrations to a volume mixing ratio of 1.5 10$^{-7}$. As a consequence, ozone is taken out of the atmosphere, if the tropopause rises to levels where the climatological values would exceed this limit. This process is reversible if the tropopause descends again. \par
Our version of the model incorporates several changes to the grid-point physics such that we obtain a more accurate representation of several physical processes in warm climates\cite{popp_initiation_2015}. The grid-point physics include representation of surface exchange, turbulence and vertical diffusion\cite{brinkop_sensitivity_1995,giorgetta_atmospheric_2013}, gravity-wave drag\cite{hines_doppler-spread_1997-1,hines_doppler-spread_1997}, radiative transfer and radiative heating\cite{mlawer_radiative_1997,iacono_radiative_2008}, convection\cite{tiedtke_comprehensive_1989,nordeng_extended_1994}, cloud cover\cite{sundqvist_condensation_1989} and cloud microphysics\cite{lohmann_design_1996}. In summary, these changes are the inclusion of the mass of water vapour when calculating the total pressure and the omission of all approximations where small specific humidities are assumed (as for example in the calculation of density). The pressure effects of water vapour are not considered for the horizontal transport. So the model is in sorts a hybrid model, with water vapour adding to the total pressure for local effects but not so for the large-scale transport. A detailed description of the modified model thermodynamics can be found in the appendix of ref. 34 \nocite{popp_climate_2014}. We will give here a short overview of the modified radiative transfer scheme as well as of the convection, cloud-cover and cloud-microphysical schemes. \par
The radiative transfer scheme has recently been described and evaluated extensively in ref. 21, but as a courtesy to the reader we will repeat some of the major features here. It is based on the Rapid Radiative Transfer Model (RRTMG)\cite{mlawer_radiative_1997,iacono_radiative_2008}, but includes some small modifications. It uses the correlated-k method to solve the radiative transfer equations in the two-stream approximation. The k-coefficients are calculated from the HITRAN (1996 and 2000) database using a line-by-line radiative transfer model (LBLRTM)\cite{mlawer_radiative_1997,iacono_radiative_2008}. The water vapour continuum is based on CKD$\_$v2.4. The shortwave radiation spectrum is divided into 14 bands, and the longwave radiation spectrum is divided into 16 bands. Since the lookup-tables of the molecular absorption coefficients are designed for a limited range of temperatures only, an exponential extrapolation for temperatures up to 400 K for the longwave radiation scheme is performed. The same extrapolation scheme is also applied to the lookup-tables for the absorption coefficients of the water vapour self-broadened continuum in the shortwave radiation scheme, but the original linear extrapolation scheme is kept for all the other absorption coefficients. The lookup-tables for the bandwise spectrally integrated Planck function and the derivative thereof with respect to temperature have been extended to 400 K. Furthermore the water vapour self-broadened continuum is introduced in the upper atmosphere radiation calculations, to account for the increase in water vapour with increasing gST. The effect of pressure broadening by water vapour on the molecular absorption coefficients is neglected, as is scattering by water vapour. The thus modified radiation scheme is not as accurate as line-by-line radiative transfer models or models using recalculated k-coefficients for the higher temperatures, but still sufficiently accurate for the task at hand as has recently been demonstrated\cite{popp_initiation_2015}. \par
ECHAM6 uses a mass-flux scheme for cumulus convection\cite{tiedtke_comprehensive_1989}, with modifications for penetrative convection to the original scheme\cite{nordeng_extended_1994}. The contribution of cumulus convection to the large-scale budgets of heat, moisture, and momentum is represented by an ensemble of clouds consisting of updrafts and downdrafts in a steady state. Depending on moisture convergence at the surface and depth of the convection cell, the model will either run in penetrative, mid-level, or shallow convection mode. The scheme allows for the formation of precipitation, but not for radiatively-active convective clouds. Instead, detrained water is passed to the cloud microphysical scheme which creates or destroys cloud condensate in a further step.     \par
ECHAM6 uses the Sundqvist scheme for fractional cloud cover\cite{sundqvist_condensation_1989}. This cloud scheme has been tuned for the use with present-day Earth's climate\cite{lohmann_design_1996,mauritsen_tuning_2012}. However, the scheme is well suited for simulations of cloud cover in warm climates as it diagnoses cloud cover directly from relative humidity, which should be crucial to cloud formation irrespective of the temperature.  \par
The cloud-microphysical scheme is described in detail in ref. 33. The scheme consists of prognostic equations for the vapour, liquid, and ice phases. There are explicit microphysics for warm-phase, mixed-phase, and ice clouds. The cloud-condensation-nuclei concentration follows a prescribed vertical profile which is typical for present-day Earth maritime conditions. Since we have no estimate of the aerosol load in a hypothetical warm climate, we assume that this profile is also a reasonable choice for warmer climates. The microphysics do not require changes for the use of the scheme in warm climates, since cloud formation in a warm-phase cloud, in which potential changes may occur, is not directly temperature-dependent (at least not in the range of temperatures we consider).  \par 
Clouds are represented in the radiative transfer calculations, assuming the so-called maximum-random-overlap assumption. Under this assumption cloud layers are assumed to be maximally overlapping if they are adjacent to one another, and randomly overlapping if they are separated by a clear layer. The absorptivity of clouds depends on their combined optical depths, the gas in which they are embedded, and the interstitial aerosol. The microphysics to determine the optical properties of the cloud particles involve the liquid-water and ice paths, cloud-drop radii, as well as liquid-water and ice content. Cloud scattering is represented as a single-scattering albedo by assuming Mie scattering from cloud droplets in the shortwave calculations, but is neglected in the longwave calculations. Clouds are not considered in the radiative transfer routines if the cloud-condensate does not exceed  10$^{-7}$ kg per kg of air.\par
In order to run the model at high temperatures, a few special settings are necessary. The time step is reduced from 2400 to 600 seconds, and the radiation time-step is reduced from 7200 to 2400 seconds, except for the simulation with a TSI of 1.15 S$_0$ where the time step is reduced to 360 and the radiation time step to 1440 seconds. Nonetheless, we sometimes encounter problems with resolved waves propagating to the top levels of the model, where their amplitude grows and where they may be reflected. In order to avoid frequent model failure due to these effects, we introduce Rayleigh friction to the vorticity and the divergence as well as increased horizontal diffusion in the top 6 layers (above around 0.75 hPa). The time constant of the Rayleigh friction is (8 d)$^{-1}$ at the sixth layer and is increased by a factor of 3.2 per layer towards the top and is hence increased by a factor of around 1000 in the top layer. The horizontal diffusion is increased by a factor of 3.2 per layer. The values of the time constant of the Rayleigh friction and the magnitude of horizontal diffusion are determined by trial and error. Since we investigate a large range of climates, it is difficult to find suitable values for these time constants, and despite these modifications the model fails occasionally. In these cases, however, the runs can be continued by slightly changing their trajectory, which is achieved by reducing the factor of multiplication per level for the Rayleigh friction from 3.2 to 3.199 for 30 days. 
\end{methods} 
%%%%%%%%%%%%%%%%%%%%%%%%%%%%%%%%%%%%%%%%%%%%%%%%%%%%%%%%%%%%%%%%%%%%%
% FIGURES
\renewcommand{\baselinestretch}{1.0} 
\clearpage
\clearpage
\clearpage
\clearpage
\clearpage
\clearpage
\clearpage
\clearpage
\renewcommand{\baselinestretch}{2.0} 

\newpage
\section*{References}
\bibliography{cloud_bib_n} 
\newpage
\end{document}